\documentclass[prl,twocolumn,showpacs,floatfix,preprintnumbers,amsmath,amssymb,superscriptaddress]{revtex4}

\usepackage{graphicx}
\usepackage{amsmath}
\usepackage{amssymb}
\usepackage{color}
\usepackage{dcolumn}
\usepackage{epsfig}
\usepackage{bm}
\usepackage{subfigure}
\usepackage[urlcolor=blue]{hyperref}
\hypersetup{backref, colorlinks=true, linkcolor=blue, citecolor=blue}

\begin{document}

\title{Superconductivity from Meissner Effect and Zero Resistivity in a Phenyl Molecule}

\author{Liu-Cheng Chen}
\affiliation{Center for High Pressure Science and Technology Advanced Research, Shanghai 201203, China}

\author{Ren-Shu Wang}
\affiliation{Center for High Pressure Science and Technology Advanced Research, Shanghai 201203, China}
\affiliation{Faculty of Materials Science and Engineering, Hubei University, Wuhan 430062, China}

\author{Jia Cheng}
\affiliation{Faculty of Materials Science and Engineering, Hubei University, Wuhan 430062, China}

\author{Xiao-Lin Wu}
\affiliation{Faculty of Physics and Electronic Technology, Hubei University, Wuhan 430062, China}

\author{Yun Gao}
\affiliation{Faculty of Materials Science and Engineering, Hubei University, Wuhan 430062, China}

\author{Zhong-Bing Huang}
\affiliation{Faculty of Physics and Electronic Technology, Hubei University, Wuhan 430062, China}

\author{Xiao-Jia Chen}
\affiliation{Center for High Pressure Science and Technology Advanced Research, Shanghai 201203, China}

\date{\today}

\begin{abstract}
Recently, phenyl molecules have been reported to exhibit Meissner effect mainly from magnetization measurements. Realizing zero-resistivity state in these materials seems a challenge due to many practical difficulties but is required to characterize the existence of superconductivity. By choosing potassium-doped tris(2-methylphenyl)bismuthine as an example, we perform temperature-dependent magnetic susceptibility and resistivity measurements at  different magnetic fields and pressures. The solid evidence for supporting superconductivity is achieved from the obtained Meissner effect and zero resistivity with the critical temperature ($T_c$) of 3.6 K at atmosphere pressure. Upon compression, we observe the gradual evolution of superconductivity from its initial phase with a parabolic behavior of $T_{c}$ to the second one with almost constant value of $T_{c}$ of 7 K. The 7 K phase seems a common feature for these newly discovered phenyl-based superconductors.
\end{abstract}


\maketitle

Exploring superconductivity with high critical temperature ($T_c$) even above room temperature in organic materials follows two pioneering works \cite{Ginzburg,little}. Superconductivity at 0.3 K was found in the quasi-one-dimensional polymer, (SN)$_x$ \cite{gree}. This served as the first effect to synthesize superconductors by using nonmetallic elements. The first quasi-one-dimensional organic superconductor named (TMTSF)$_2$PF$_6$ \cite{jerom} was found by the application of pressure. Since then, many organic superconducting families were discovered, including the charge transfer complexes \cite{Ishiguro}, fullerides \cite{Hebard,gani}, graphites \cite{Emery,jskim}, and polycyclic aromatic hydrocarbons \cite{Mitsu,Wang,Xue}. Although these superconductors possess many novel physical properties, their $T_c$'s are still bound below 40 K. Recently, $p$-oligophenyls are being examined as a new class of superconductors following a series of works of Wang $et$ $al$ \cite{wang1,wang2,wang3}. As the models of polyparaphenylene $-$ one type of conducting polymers, these materials are chain compounds with benzene ring as their basic unit and connected by single C-C bond in $para$ position. It has long been recognized \cite{3,4,5,6,7,8,20,Zhong} that doping $p$-oligophenyls and conducting polymers results in the formation of polarons and bipolarons with the latter as the stable sate. Bipolarons have ever been proposed to account for the electrical transport in conducting polymeric materials \cite{21}. Bipolarons can be considered as localized spatially non-overlapping Cooper pairs, analogous to the one in the BCS theory of superconductivity. Since the electron pairing already takes place and stabilizes at room temperature, high-temperature superconductivity would be expected in such systems if long-range phase coherence can be induced and enhanced through some methods such as chemical doping of pressure \cite{emery}. This may be the key reason why superconductivity can occur above 120 K in $p$-terphenyl \cite{wang3} and $p$-quaterphenyl \cite{jiafeng}. A superconducting phase with \emph{T$_c$} as high as 107 K was also reported in K-doped \emph{p}-terphenyl flake \cite{Neha}. Superconductivity-like behaviors above 120 K were observed in $p$-terphenyl and $p$-quaterphenyl as well \cite{wen}. An energy gap persisting up to 120 K was reported in surface K-doped \emph{p}-terphenyl crystals from the photoemission spectroscopy \cite{Li}. The similar feature without obvious changes by the applied magnetic fields \cite{ren} indicates high upper critical field.

\begin{figure*}
\begin{center}
\includegraphics[width=0.98\textwidth]{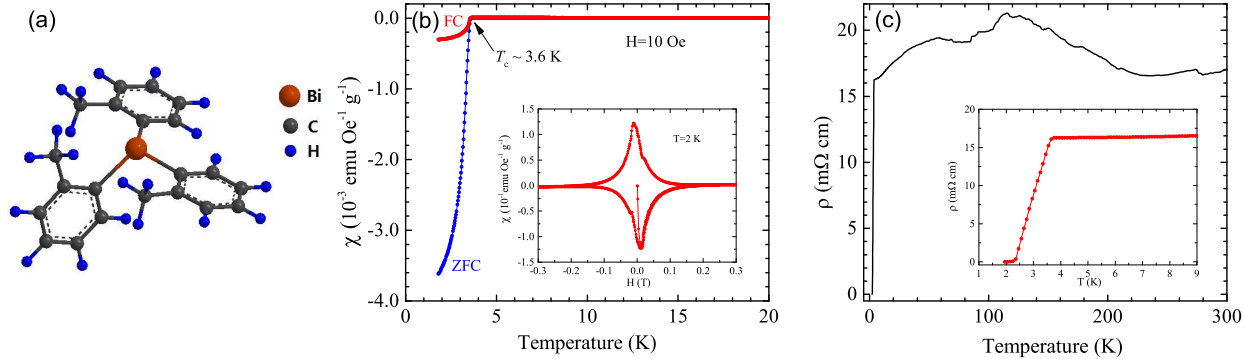}
\end{center}
\caption{Evidence for superconductivity in potassium-doped tris(2-methylphenyl)bismuthine at ambient pressure. (a) Structural formula and atomic arrangement of tris(2-methylphenyl)bismuthine. (b) Temperature dependence of the $dc$ magnetic susceptibility for potassium-doped tris(2-methylphenyl)bismuthine measured in zero-field cooling (ZFC) and field-cooling (FC) runs at a low magnetic field of 10 Oe. The inset shows the magnetic hysteresis loop with applied field up to 0.3 T at the temperature of 2 K. (c) Temperature dependence of the resistivity of potassium-doped tris(2-methylphenyl)bismuthine at ambient pressure. The inset displays the enlarged view of the low temperature resistivity.}
\end{figure*}

Meissner effect has been already reported from the $dc$ magnetic susceptibility ($\chi$) measurements in this family upon potassium doping including biphenyl \cite{ZhangK}, $p$-terphenyl \cite{wang1,wang2,wang3}, $p$-quaterphenyl \cite{jiafeng}, and $p$-quinquephenyl \cite{gehuang}. The effect was established \cite{wang1,wang2,wang3,ZhangK,jiafeng,gehuang} from the typical shape of $\chi$ in both the zero-field cooling (ZFC) and field-cooling (FC) runs, the systematic shift of the ZFC $\chi$ downward to low temperatures by the applied magnetic field till saturation at a certain critical field, and the reduction of the magnetization with the application of magnetic field up to a certain critical value together with a typical diamond shape afterward. The $ac$ $\chi$ also gives a similar shape for its real part as the ZFC $\chi$ curve. The observed peak and the almost zero signal upon further cooling in the imaginary part of the $ac$ $\chi$ curve \cite{wang1,ZhangK,gehuang} indicate the realization of zero-resistivity in the superconducting state \cite{hein,gomo}. However, the direct resistance measurements just showed the downshift of the temperature - resistance curve with the applied magnetic field and the saturation at a certain critical field \cite{ZhangK}. The zero resistivity as one of the two fundamental characters of superconductivity has not been reached yet in this family.

To pin down this issue, here we choose potassium-doped tris(2-methylphenyl)bismuthine to examine the possibility of superconductivity from both magnetization and resistivity measurements. It has three phenyl rings attaching to a bismuth atom with joint three methyl groups at the ortho site of each benzene ring, as indicated in Fig. 1(a). This phenyl molecule is a derivative of triphenylbismuth, which was reported to show superconductivity at 3.5 K and/or 7.2 K upon potassium doping \cite{renshu}. We hope to identify superconductivity in this compound from the observations of the Meissner effect and zero resistivity. The obtained evidence of the zero resistivity is expected to provide the foundation for the existence of superconductivity in alkali metal doped phenyl-based materials.

Sample synthesis of potassium-doped tris(2-methylphenyl)bismuthine is the same method as that of intercalated triphenylbismuth \cite{renshu}. Due to the sensitivity of the samples to water and oxygen , all the experiments were prepared in a glove box with O$_2$ and H$_2$O levels less than 0.1 ppm. The magnetization measurements were completed in a Magnetic Properties Measurement System (Quantum Design MPMS3). The resistivity at ambient pressure and high pressures were measured in a nonmagnetic diamond anvil cell (DAC) made of Cu-Be alloy with the help of standard four-probe method. Two symmetrical anvil culets with the diameter of 500 $\mu$m were fixed in the customized cell which could give a high pressure environment. The sample chamber with the diameter of 200 $\mu$m was formed with an insulated gasket. Four Pt wires were fixed on the upper diamond with the standard four-probe geometry. Then, four external Cu wires were linked to the four Pt wires using for the resistivity measurements in Physical Properties Measurement System (PPMS). The sample was filled in the sample chamber for several times in order to overcover the chamber in the glove box. Finally, the DAC was closed after putting a ruby in the sample chamber. For the resistivity at ambient pressure, the samples were pelletized for better contacting with the electrodes. Pressure was calibrated by using the ruby fluorescence shift \cite{maoh} at room temperature. 

\begin{figure}[tbp]
\begin{center}
\includegraphics[width=0.45\textwidth]{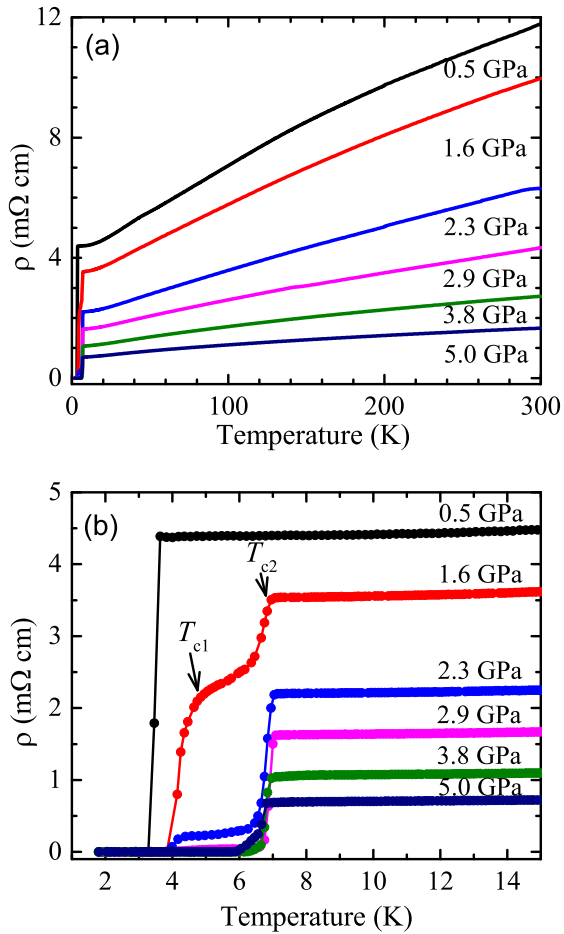}
\end{center}
\caption{(a) Temperature dependence of the resistivity of potassium-doped tris(2-methylphenyl)bismuthine at virious pressures up to 5.0 GPa. (b) The enlarged view of the low-temperature resistivities. $T_{c1}$ and $T_{c2}$ are the critical temperatures for the first and second superconducting phases, respectively.}
\end{figure}

Figure 1 summarizes our results of the magnetic susceptibility and resistivity measurements on potassium-doped tris(2-methylphenyl)bismuthine at ambient pressure. From the temperature dependence of the $dc$ magnetic susceptibility ($\chi$) with zero-field cooling (ZFC) and field cooling (FC) runs at 10 Oe [Fig. 1(b)], one can see a sudden drop in $\chi$ around 3.6 K in both the ZFC and FC runs. The sudden drop of $\chi$, originating from the well-defined Meissner effect, manifests the occurrence of superconductivity in this sample. The $T_c$ is defined as the temperature where the sharp drop takes place. The shielding volume fraction extracted from $\chi$ at 1.8 K is only about 6.8\%. The inset of Fig. 1(b) shows the magnetic hysteresis loop with scanning magnetic field up to 0.3 T at the temperature of 2 K. This clear diamond-like hysteresis loop again supports the Meissner effect in this material and also provides electrical transport evidence for this sample as a typical type-II superconductor.

Figure 1(c) displays electrical transport measurement of potassium-doped tris(2-methylphenyl)bismuthine at ambient pressure. It can be seen that the change of the resistivity with temperature is not regular. There seemingly exists a hump at around 120 K. Above that, the resistivity shows non-metallic feature. It changes to metallic-like behavior at low temperatures. This phenomenon is probably due to the weak linkage inside the sample or the poor contact between the sample and electrode. However, as temperature is decreased, the resistivity suddenly shows a sharp drop and then gets to zero at a certain temperature. The zero-resistivity is the character of superconductivity. This observation serves the first evidence for supporting the existence of superconductivity in a phenyl molecule. The temperature, below which a sudden drop of the resistivity is observed, is just the critical temperature for this new superconductor. Its value is exactly the same as that detected from the magnetic susceptibility measurements [Fig. 1(b)]. The inset of Fig. 1(c) shows the enlarged view of the low temperature resistivity. The zero-resistance behavior with $T_c$ $\sim$ 3.6 K can be clearly observed. Thus, potassium-doped tris(2-methylphenyl)bismuthine is identified to be a new superconductor from both the Meissner effect and zero-resistivity state.

\begin{figure}[tbp]
\includegraphics[width=0.45\textwidth]{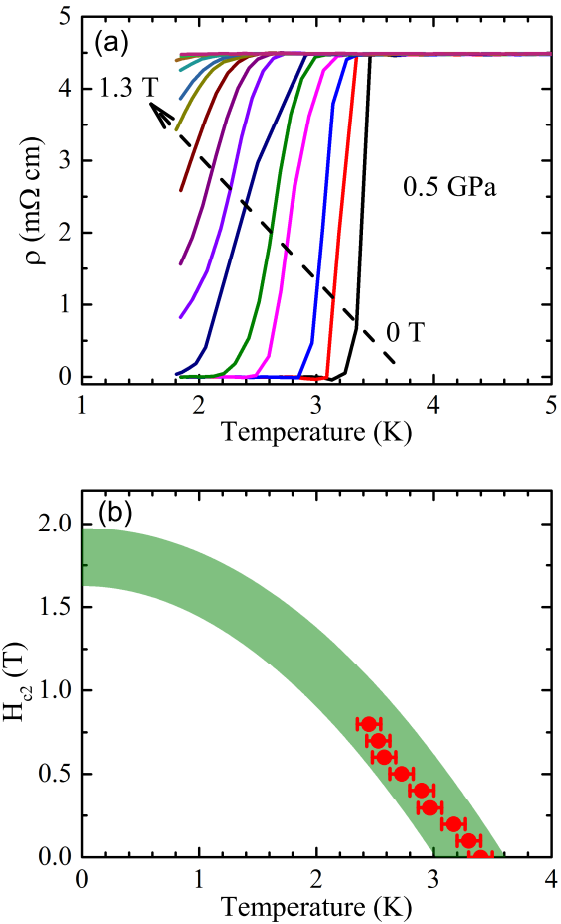}
\caption{(a) Temperature-dependent resistivity at various magnetic fields from 0.0 to 1.3 Tesla at pressure of 0.5 GPa. (b) Upper critical field ($H_{c2}$) as a function of temperature. The color area represents the calculated $H_{c2}$ from the Werthamer-helfand-Hohenberg equation.}
\end{figure}

It is generally believed \cite{rmp} that the superconducting shielding volume fraction would be increased and $T_{c}$ would be enhanced when the atoms get closer in the unit cell. Pressure has been recognized to be an effective means in inducing superconductivity for highly compressible molecular materials \cite{jerom,gani,tani}. We examined this idea on this newly discovered superconductor by performing high-pressure resistivity measurements. The results are shown in Fig. 2. As can be seen, the temperature-resistivity curve is systematically moved downward with increasing pressures. The metallic normal state is clearly seen for this new superconductor at high pressures. The similar metallic normal state has also been observed in another phenyl-based superconductor \cite{ZhangK}. However, the edge shared K-doped picene possesses non-metallic normal state, though zero-resistivity was also observed \cite{Teranishi}. The increased conductivity is apparently observed by the applied pressure. The enhanced $T_{c}$ is also found upon compression.

Interestingly, the gradual evolution of superconducting phase with pressure is observed in this phenyl molecule. There exist two superconducting phases. From the enlarged view of the temperature-resistivity curve at low temperatures [Fig. 2(b)], one can see a sudden drop of the resistivity and the zero-resistivity behavior at 0.5 GPa. A new superconducting phase emerges when pressure is increased to 1.6 GPa. Two transitions with sudden resistivity drops signal the superconducting critical temperatures at around 3.8 K ($T_{c1}$) and 7 K ($T_{c2}$) for the first and second phases, respectively. Upon further compression, the first superconducting phase is gradually suppressed till disappearing up to 3 GPa and the second superconducting phase becomes the dominant one.

The obtained superconductivity of potassium-doped tris(2-methylphenyl)bismuthine is further confirmed by the resistivity measurements with applied magnetic fields. Figure 3(a) shows temperature dependence of the resistivity at pressure of 0.5 GPa and at various magnetic fields up to 1.3 T. The suppression of superconductivity can be found by the application of magnetic fields. The temperature-dependent resistivity curve systematicaly shifts toward lower temperatures with increasing magnetic fields. Meanwhile, the temperature span of superconducting transition becomes significant broad as field is increased. Superconductivity is completely destroyed at the magnetic field of 1.3 Tesla in the studied temperature range. The suppression of superconductivity by the applied magnetic fields has been observed in many phenyl-based molecules from the magnetic susceptibility measurements \cite{wang1,wang2,wang3,jiafeng,gehuang,renshu}. The similar behavior observed here supports the existence of superconductivity in these systems. The temperature-dependent resistivity curves at various magnetic fields allow the determination of an important superconducting parameter $-$ the upper critical field ($H_{c2}$). $H_{c2}$ is defined by using the onset $T_c$ criteria, which is determined by the first dropped point deviated from the linear resistivity curve. Figure 3(b) summarizes the temperature dependence of $H_{c2}$. Based on the Werthamer-Helfand-Hohenberg equation \cite{BCS}: $H_{c2}(0) = 0.693[-(dH_{c2}/dT)]_{T_c}T_c$, one can obtain the value of $H_{c2}(0)$. The calculated $H_{c2}(0)$ is about 1.8 Tesla at 0 K. The colorful area is extrapolated by using the formula of $H_{c2}(T) = H_{c2}(0)[1-(T/T_c)^2]/[1+(T/T_c)^2]$ based on the Ginzburg-Landau theory.

\begin{figure}[tbp]
\includegraphics[width=0.48\textwidth]{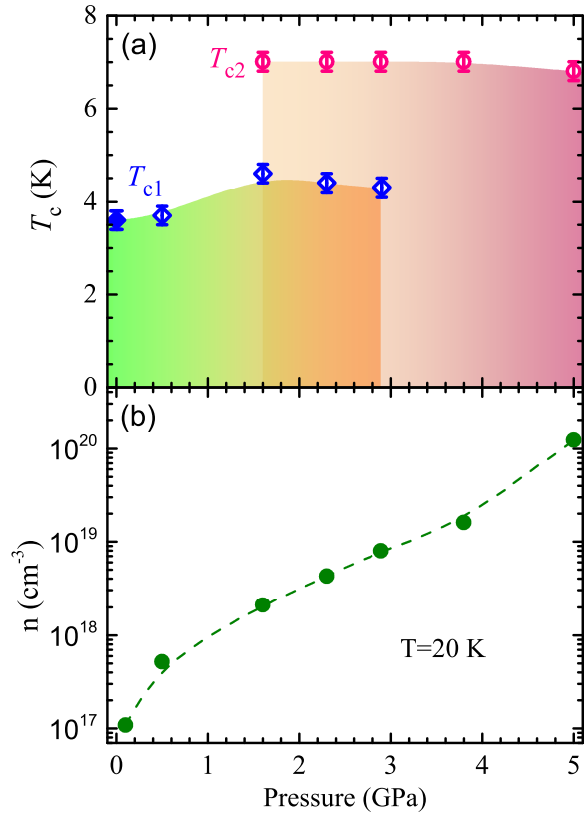}
\caption{(a) Pressure-temperature phase diagram of potassium-doped tris(2-methylphenyl)bismuthine. The intermediate region represents the coexistence of the two superconducting states. (b) Pressure dependence of the carrier concentration calculated from the Hall coefficient measurements taken at 20 K.}
\end{figure}

From both the magnetic susceptibility and resistivity measurements, one can establish the superconducting phase diagram for potassium-doped tris(2-methylphenyl)bismuthine as a function of pressure [Fig. 4(a)]. It can be seen that the initial superconducting phase maintains at pressures up to 2.9 GPa. $T_{c1}$ first increases with increasing pressure and then declines after passing a maximum. As pressure is increased to 1.6 GPa, the second superconducting phase appears and $T_{c2}$ keeps almost a constant at high pressures. This behavior indicates that the second superconducting phase is not sensitive to the applied pressure. Notably, the superconducting phase with $T_c$ about 7 K has been generally observed in phenyl-based molecules \cite{wang1,wang2,wang3,jiafeng,gehuang,renshu} as well as the edge shared aromatic hydrocarbons \cite{Mitsu,Wang,Xue}. Therefore, the common superconducting phase with $T_c$ $\sim$ 7 K may be the uniform character of phenyl-based compounds.

Hall coefficient measurements were performed at temperature of 20 K and at various pressures. The obtained carrier concentration ($n$) as a function of pressure is shown in Fig. 4(b). The $n$ is calculated through the formula $n$=1/$eR_H$, where $e$ is the unit of the charge. It can be seen that $n$ has an obvious increase in the whole pressure range, from 10$^{17}$ to 10$^{20}$ cm$^{-3}$, which is very beneficial to superconductivity in this sample. The emergence of the second superconducting phase is accompanied by the increased carrier concentration. This phenomenon indicates the important role of the carrier concentration played for superconductivity. This provides a clue to the theory developments for understanding superconductivity in these phenyl-based molecules \cite{Mazziotti,Geilhufe}.

In summary, we have conducted both the magnetic susceptibility and resistivity measurements on a phenyl molecule $-$ potassium-doped tris(2-methylphenyl)bismuthine. Superconductivity with $T_c$ of 3.6 K is firmly identified from the obtained Meissner effect and zero-resistivity state for this material at atmosphere pressure. With the application of pressures, $T_{c}$ first exhibits a parabolic-like shape and disappears at pressure of 2.9 GPa for the first superconducting phase. The second superconducting phase with $T_c$ of 7.2 K begins to appear at pressure of 1.6 GPa, and then keeps a constant value upon further compression. These findings offer a solid foundation for the existence of superconductivity in phenyl-based molecules.


\begin{references}


\bibitem{Ginzburg} V. L. Ginzburg, High-temperature superconductivity dream or reality? Sov. Phys. Usp. \textbf{19}, 174 (1976).

\bibitem{little} W. A. Little, Possibility of synthesizing an organic superconductor. Phys. Rev. \textbf{134}, A1416 (1964).

\bibitem{gree} R. L. Greene, G. B. Street, and L. J. Suter, Superconductivity in polysulfur nitride (SN)$_x$. Phys. Rev. Lett. \textbf{34}, 577 (1975).

\bibitem{jerom} D. J$\acute{e}$rome, A. Mazaud, M. Ribault, and K. Bechgaard, Superconductivity in a synthetic organic conductor (TMTSF)$_2$PF$_6$. J. Phys. Lett. \textbf{41}, 95 (1980).

\bibitem{Ishiguro} T. Ishiguro, K. Yamaji, and G. Saito, {\it Organic Superconductors} (2nd edition, Springer-Verlag, 1997).

\bibitem{Hebard} A. F. Hebard, M. J. Rosseinsky, R. C. Haddon, D. W. Murphy, S. H. Glarum, T. T.M. Palstra, A. P. Ramirez, and A. R. Kortan, Potassium-doped C$_{60}$. Nature \textbf{350}, 600 (1991).

\bibitem{gani} A. Y. Ganin, Y. Takabayashi, Y. Z. Khimyak, S. Margadonna, A. Tamai, M. J. Rosseinsky, and K. Prassides, Bulk superconductivity at 38 K in a molecular system. Nat. Mater. \textbf{7}, 367 (2008).

\bibitem{Emery} N. Emery, C. H$\acute{e}$rold, M. d'Astuto, V. Garcia, Ch. Bellin, J. F. Mareche, P. Lagrange, and G. Loupias, Superconductivity of bulk CaC$_6$. Phys. Rev. Lett. \textbf{95}, 087003 (2005).

\bibitem{jskim} J. S. Kim, L. Boeri, J. R. O'Brien, F. S. Razavi, and R. K. Kremer, Superconductivity in heavy alkaline-earth intercalated graphites. Phys. Rev. Lett. \textbf{99}, 027001 (2007).

\bibitem{Mitsu} R. Mitsuhashi, Y. Suzuki, Y. Yamanari, H. Mitamura, T. Kambe, N. Ikeda, H. Okamoto, A. Fujiwara, M. Yamaji, N. Kawasaki, Y. Maniwa, and Y. Kubozono, Superconductivity in alkali-metal-doped picene. Nature \textbf{464}, 76 (2010).

\bibitem{Wang} X. F. Wang, R. H. Liu, Z. Gui, Y. L. Xie, Y. J. Yan, J. J. Ying, X. G. Luo, and X. H. Chen, Superconductivity at 5 K in alkali-metal-doped phenanthrene. Nat. Commun. \textbf{2}, 507 (2011).

\bibitem{Xue} M. Xue, T. Cao, D. Wang, Y. Wu, H. Yang, X. Dong, and G. F. Chen, Superconductivity above 30 K in alkali-metal-doped hydrocarbon. Sci. Rep. \textbf{2}, 389 (2012).

\bibitem{wang1} R. S. Wang, Y. Gao, Z. B. Huang, and X. J. Chen, Superconductivity in \emph{p}-terphenyl. arXiv:1703.05803 (2017).

\bibitem{wang2} R. S. Wang, Y. Gao, Z. B. Huang, and X. J. Chen, Superconductivity at 43 K in a single C-C bond linked terphenyl. arXiv:1703.05804 (2017).

\bibitem{wang3} R. S. Wang, Y. Gao, Z. B. Huang, and X. J. Chen, Superconductivity above 120 kelvin in a chain link molecule. arXiv:1703.06641 (2017).

\bibitem{3} J. L. Br$\acute{e}$das, R. R. Chance, and R. Silbey, Comparative theoretical study of the doping of conjugated polymers: Polarons in polyacetylene and polyparaphenylene. Phys. Rev. B \textbf{26}, 5843 (1982).

\bibitem{4} J. L. Br$\acute{e}$das, B. Th$\acute{e}$mans, J. G. Fripiat, J. M. Andr$\acute{e}$, and R. R. Chance, Highly conducting polyparaphenylene, polypyrrole, and polythiophene chains: An ab initio study of the geometry and electronic-structure modifications upon doping. Phys. Rev. B \textbf{29}, 6761 (1984).

\bibitem{5} R. R. Chance, J. L. Br$\acute{e}$das, and R. Silbey, Bipolaron transport in doped conjugated polymers. Phys. Rev. B \textbf{29}, 4491 (1984).

\bibitem{6} J. C. Scott, P. Pfluger, M. T. Krounbi, and G. B. Street, Electron-spin-resonance studies of pyrrole polymers: Evidence for bipolarons. Phys. Rev. B \textbf{28}, 2140 (1983).

\bibitem{7} J. H. Kaufman, N. Colaneri, J. C. Scott, and G. B. Street, Evolution of polaron states into bipolarons in polypyrrole. Phys. Rev. Lett. \textbf{53}, 1005 (1984).

\bibitem{8} G. Crecelius, M. Stamm, J. Fink, and J. J. Ritsko, AsF$_5$-doped polyparaphenylene: Evidence for polaron and bipolaron formation. Phys. Rev. Lett. \textbf{50}, 1498 (1983).

\bibitem{20} Y. Furukawa, H. Ohtsuka, and M. Tasumi, Raman studies of intact and sodium doped $^{13}$C-substituted poly-$p$-phenylene. J. Raman Spectrosc. \textbf{24}, 551 (1993).

\bibitem{Zhong} G. H. Zhong, X. H. Wang, R. S. Wang, J. X. Han, C. Zhang, X. J. Chen, and H. Q. Lin, Structural and bonding character of potassium-doped $p$-terphenyl superconductors. J. Phys. Chem. C. \textbf{122}, 3801 (2017).

\bibitem{21} C. K. Chiang, C. R. Fincher, Jr., Y. W. Park, A. J. Heeger, H. Shirakawa, E. J. Louis, S. C. Gau, and A. G. MacDiarmid, Electrical conductivity in doped polyacetylene. Phys. Rev. Lett. \textbf{39}, 1098 (1977).

\bibitem{emery} V. J. Emery and S. A. Kivelson, Importance of phase fluctuations in superconductors with small superfluid density. Nature \textbf{374}, 434 (1995).

\bibitem{jiafeng} J. F. Yan, R. S. Wang, K. Zhang, and X. J. Chen, Observation of Meissner effect in potassium-doped $p$-quaterphenyl. arXiv:1801.08220 (2018).

\bibitem{Neha} P. Neha, V. Sahu, and S. Patnaik, Facile synthesis of potassium intercalated $p$-terphenyl and signatures of a possible high $T_c$ phase. arXiv:1712.01766 (2017).

\bibitem{wen} W. H. Liu, H. Lin, R. Z. Kang, X. Y. Zhu, Y. Zhang, S. X. Zheng, and H. H. Wen, Magnetization of potassium doped $p$-terphenyl and $p$-quaterphenyl by high pressure synthesis. Phys. Rev. B \textbf{96}, 224501 (2017).

\bibitem{Li} H. X. Li, X. Q. Zhou, S. Parham, T. Nummy, J. Griffith, K. Gordon, E. L. Chronister, and D. S. Dessau, Spectroscopic evidence of low energy gaps persisting towards 120 Kelvin in surface-doped $p$-terphenyl crystals. arXiv:1704.04230v2 (2017).

\bibitem{ren} M. Q. Ren, W. Chen, Q. Liu, C. Chen, Y. J. Qiao, Y. J. Chen, G. Zhou, T. Zhang, Y. J. Yan, and D. L. Feng, Observation of novel gapped phases in potassium doped single layer \emph{p}-terphenyl on Au (111). arXiv preprint arXiv:1705.09901 (2017).

\bibitem{ZhangK} K. Zhang, R. S. Wang, A. J. Qin, and X. J. Chen, Superconductivity in potassium doped 2,2$'$-bipyridine. arXiv:1801.06320 (2018).

\bibitem{gehuang} G. Huang, R. S. Wang, and X. J. Chen, Observation of Meissner effect in potassium-doped $p$-quinquephenyl. arXiv:1801.06324 (2018).

\bibitem{hein} R. A. Hein, \emph{ac} magnetic susceptibility, Meissner effect, and bulk superconductivity. Phys. Rev. B \textbf{33}, 7539 (1986).

\bibitem{gomo} F. G$\ddot{o}$m$\ddot{o}$ry, Characterization of high-temperature superconductors by \emph{ac} susceptibility measurements. Supercond. Sci. Tech. \textbf{10}, 523 (1997).

\bibitem{renshu} R. S. Wang, J. Cheng, X. L. Wu, H. Yang, X. J. Chen, Y. Gao, and Z. B. Huang, Highly reproducible superconductivity in potassium-doped triphenylbismuth. arXiv:1802.03320vl.

\bibitem{maoh} H. K. Mao, P. M. Bell, J. W. Shaner, and D. J. Stembey, Specific volume measurements of Cu, Mo, Pd, and Ag and calibration of the ruby R$_1$ fluorescence pressure gauge from 0.06 to 1 Mbar. J. Appl. Phys. \textbf{49}, 3276 (1978).

\bibitem{rmp} H. K. Mao, X. J. Chen, Y. Ding, B. Li and L. Wang, Solids, liquids, and gases under high pressure. Rev. Mod. Phys. \textbf{90}, 015007 (2018).

\bibitem{tani} H. Taniguchi, M. Miyashita, K. Uchiyama, K. Satoh, N. M$\hat{o}$ri, H. Okamoto, K. Miyagawa, K. Kanoda, M. Hedo, and Y. Uwatoko, Superconductivity at 14.2 K in layered organics under extreme pressure. J. Phys. Soc. Jpn. \textbf{72}, 468 (2003).

\bibitem{Teranishi} K. Teranishi, X. X. He, Y. Sakai, M. Izumi, H. Goto, R. Eguchi, Y. Takabayashi, T. Kambe, and Y. Kubozono, Observation of zero resistivity in K-doped picene. Phys. Rev. B \textbf{87}, 060505(R) (2013).

\bibitem{BCS} N. R. Werthamer, E. Helfand, and P. C. Hohenberg, Temperature and purity dependence of the superconducting critical field, H$_{c2}$. III. electron spin and spin-orbit effects. Phys. Rev. \textbf{147}, 295 (1966).

\bibitem{Mazziotti} M. V. Mazziotti, A. Valletta, G. Campi, D. Innocenti, A. Perali, and A. Bianconi, Possible fano resonance for high $T_c$ multi-gap superconductivity in $p$-terphenyl doped by K at the Lifshitz transition. Europhys. Lett. \textbf{118}, 37003 (2017).

\bibitem{Geilhufe} R. M. Geilhufe, S. S. Borysov, D. Kalpakchi, and A. V. Balatsky, Towards novel organic high-$T_c$ superconductors: Data mining using density of states similarity search. Phys. Rev. Mater. \textbf{2}, 2475 (2018).


\end{references}
\end{document}